\documentclass[prd,aps,nofootinbib,onecolumn,showpacs,12pt]{revtex4-1}
\usepackage{graphicx,epsfig}

\usepackage[table]{xcolor}
\usepackage{colortbl}
\definecolor{lightgray}{gray}{0.9}

\usepackage{epsf}
\usepackage{graphicx}

\begin{document}

\title{  \bf Semileptonic transition of P wave bottomonium  $\chi_{b0}(1P)$ to $B_{c}$ meson}
\author{ K. Azizi$^{1,\dag}$ ,  H. Sundu$^{2,\ddag}$,  J. Y. S\"{u}ng\"{u}$^{2,*}$
\\$^1$Department of Physics, Dogus University, Ac{\i}badem-Kad{\i}k\"oy, 34722 Istanbul, Turkey\\
$^2$Department of Physics, Kocaeli University, 41380 Izmit, Turkey\\
$^\dag$e-mail:kazizi@dogus.edu.tr \\
$^\ddag$e-mail:hayriye.sundu@kocaeli.edu.tr \\
$^*$e-mail:jyilmazkaya@kocaeli.edu.tr}

\begin{abstract}

Taking into account the two-gluon condensate contributions, the transition form factors enrolled to the low energy effective Hamiltonian describing the  semileptonic $\chi_{b0}\rightarrow
B_{c}\ell\overline{\nu}, (\ell=(e,\mu,\tau))$ decay channel are
calculated within three-point QCD sum rules. The fit function of the form factors then are used to estimate the  decay width of the decay mode under consideration.

\end{abstract}

\pacs{11.55.Hx, 14.40.Pq,  13.20.-v, 13.20.Gd}

\maketitle

%%%%%%%%%%%%%%%%%%%%%%%%%%%%%%%%%%%%%%%%%%%%%%%%%%%%%%%%%%%%%%%%%%%%
%\section{Introduction}
%%%%%%%%%%%%%%%%%%%%%%%%%%%%%%%%%%%%%%%%%%%%%%%%%%%%%%%%%%%%%%%%%%%%

\section{Introduction}

The quarkonia, especially the bottomonium $b \bar b$ states, are approximately non-relativistic systems since they do not contain  intrinsically relativistic light quarks. Hence, these states are the best
candidates to examine the hadronic dynamics and investigate both perturbative and non-perturbative characteristic  of QCD. In the past, mainly theoretical calculations on the properties of these states had 
been made  using potential model or its extensions like the Coulomb gauge model (see for instance \cite{Godfrey-y,Ebert-y,Crater-y,Wang-y,Dudek-y,Guo-y} and references therein). In \cite{Lucha}, 
 both potential model and QCD sum rule approach have been applied to extract the ground-state decay constant of mesons containing heavy b quark. It is stated  that  the QCD sum rule technique
gives  more reliable and accurate determination of bound-state characteristics compared to the potential models by tunning the continuum threshold parameter. The QCD sum rule approach \cite{shifman} is one of
 the most powerful and applicable tools to hadron physics. This model has been widely applied to investigate the spectroscopy of hadrons and their electromagnetic, weak and strong decays. 
The obtained results have very good consistencies with the experimental data to date within the typical (10-20)\% error bars of the technique.

The present work is dedicated to investigation of the semileptonic transition of scalar P wave bottomonium  $\chi_{b0}(1P)$ meson with quantum numbers $I^G(J^{P C }) = 0^+ (0^{ +  + } )$ into the pseudoscalar
 $B_c$ meson. The   $\chi_{b0}(1P)$ state has been observed first
 in radiative decay of the $\Upsilon(2S)$ \cite{Nakamura} and recently has been confirmed by ATLAS Collaboration \cite{atlas} together with the higher $\chi_{b}(2P)$ and $\chi_{b}(3P)$ states. In the latter, these 
quarkonia states have been produced  in proton-proton collisions at the Large Hadron Collider (LHC) at $\sqrt{s} = 7~ TeV$ and through their radiative decays to $\Upsilon(1S,2S)$ 
with $\Upsilon\rightarrow \mu^+\mu^-$. Our previous theoretical results \cite{E Veliev} on the mass of these states done both in
 vacuum and finite temperature QCD  are in good agreement with the experimental results   \cite{atlas}. Note that, we also have applied the QCD sum rules approach both in vacuum and finite temperature to investigate the spectroscopy of the pseudoscalar, vector and tensor 
quarkonia in \cite{kazem1,kazem2,kazem3}. As we know the masses and decay constants of the quarkonia, it is possible to investigate their electromagnetic, weak (leptonic-semileptonic) and strong
decays. Considering such decay channels can help us obtain more information about the nature of the scalar  $\chi_{b0}(1P)$  meson as well as  
perturbative and non-perturbative aspects of QCD.

The layout of this article is as follows. In the next section, we derive the QCD sum rules for the form factors appearing in the amplitude of
 the semileptonic decay channel under consideration. To do so, we take into account the two-gluon condensates as the non-perturbative contributions to the correlation function.
Section III is devoted to our numerical analysis of the obtained form factors and their behavior in terms of the transferred momentum squared. In this section, we also numerically
 estimate the decay width of the semileptonic $\chi_{b0}\rightarrow
B_{c}\ell\overline{\nu}$ decay mode. The last section encompasses our concluding remarks.

%%%
%%%
\section{QCD Sum Rules for  Transition Form Factors of $\chi_{b0}\rightarrow B_c\ell\overline{\nu}$ }
The hadronic event under consideration can be described in terms of quark degrees of freedom by the process $b\rightarrow cl\bar{\nu}$ at tree-level, whose effective Hamiltonian can be written as:
\begin{equation}\label{Heff}
{\cal H}_{eff}=\frac{G_{F}}{\sqrt{2}} V_{cb}~\overline{\nu}
~\gamma_{\mu}(1-\gamma_{5})l~\overline{c}
~\gamma_{\mu}(1-\gamma_{5}) b,
\end{equation}
where $G_F$ is the Fermi weak coupling constant and $V_{cb}$ is an element of the Cabibbo-Kobayashi-Maskawa (CKM)
mixing matrix. 
The transition amplitude is obtained via
\begin{equation}
 {\cal M}=\langle B_c(p')\mid {\cal H}_{eff}\mid\chi_{b0}(p)\rangle,
\end{equation}
or
\begin{equation}\label{Matrix element}
{\cal M}=\frac{G_{F}}{\sqrt{2}} V_{cb}\overline{\nu}
\gamma_{\mu}(1-\gamma_{5})l\langle B_c(p')\mid\overline{c}
\gamma_{\mu}(1-\gamma_{5})b\mid\chi_{b0}(p)\rangle.
\end{equation}
To proceed, we need to know the transition matrix element $\langle B_c(p')\mid\overline{c}
\gamma_{\mu}(1-\gamma_{5})b\mid\chi_{b0}(p)\rangle$ whose vector part do not
contribute due to parity considerations, i.e.,
\begin{equation}\label{current}
\langle B_c(p')|\overline{c}\gamma_\mu b|\chi_{b0}(p)\rangle=0.
\end{equation}
The axial-vector part of transition matrix element can be parameterized in terms of  form factors  as
\begin{equation}\label{f1Pmuf2qmu}
\langle B_c(p')\mid\overline{c}\gamma_{\mu}\gamma_5 b\mid
\chi_{b0}(p)\rangle=f_{1}(q^{2})P_{\mu}+f_{2}(q^{2})q_{\mu},
\end{equation}
where $f_{1}(q^2)$ and $f_{2}(q^2)$ are  transition form factors; and
$P_{\mu}=(p+p')_{\mu}$ and $q_{\mu}=(p-p')_{\mu}$.

Our main goal in the present section is to calculate the transition form factors applying the QCD sum rules technique. 
The starting point is to consider the following tree-point correlation function as the main
ingredient of the model:
\begin{eqnarray}\label{cor.fun}
\Pi _{\mu}=i^2\int d^{4}x\int d^4ye^{-ipx}e^{ip'y}\langle0 \mid
{\cal T}\Big\{J_{B_c}(y) J_{\mu}^{A;V}(0) J^\dag_{\chi_{b0}}(x)\Big\}\mid 0\rangle,
\end{eqnarray}
where ${\cal T}$ is the time ordering product,
$J_{B_c}(y)=\overline{c}\gamma_{5}b$ and
$J_{\chi_{b0}}(x)=\overline{b}Ub$ are the interpolating currents of
the $B_c$ and $\chi_{b0}$ mesons, respectively; and
$J_{\mu}^{V}(0)=~\overline{c}\gamma_{\mu}b $ and
$J_{\mu}^{A}(0)=~\overline{c}\gamma_{\mu}\gamma_{5}b$ are the vector
and axial-vector parts of the transition current. Following the general idea in the QCD sum rules technique, we calculate this correlation function once in terms of hadronic degrees of freedom called physical or phenomenological side and
the second in terms of QCD degrees of freedom (quarks and gluons and their interaction with QCD vacuum) called the QCD side. The latter is done  in the deep Euclidean region by the help of operator product expansion (OPE).
These two representations are then matched together, using the quark-hadron duality assumption, through a double dispersion relation to obtain the QCD sum rules for the form factors. As we deal with the ground states in this approach, we shall separate 
the ground state from
the higher states and continuum. This is done by two mathematical operations called Borel transformation and continuum subtraction. Such transformations bring some auxiliary parameters namely two Borel mass parameters
and two continuum thresholds for which we will find their working regions in the next section.

The phenomenological side of the correlation function is obtained inserting  two complete sets of intermediate
states with the same quantum numbers as the interpolating currents
$J_{B_c}$ and $J_{\chi_{b0}}$.  As a result, we obtain
\begin{eqnarray} \label{int.states}
\Pi^{PHYS}_{\mu}&=&\frac{\langle0\mid J_{B_c}(0)\mid
B_c(p')\rangle\langle B_c(p')\mid J_{\mu}^{A}(0)\mid
\chi_{b0}(p)\rangle\langle\chi_{b0}(p)\mid J^\dag_{\chi_{b0}}(0)\mid
0\rangle}{(p'^2-m_{B_c}^2)(p^2-m_{\chi_{b0}}^2)}+\cdots,
\end{eqnarray}
where $\cdots$ represents the contributions coming from higher
states and continuum. Besides the transition matrix elements defined previously,  the matrix elements of interpolating current between the vacuum and hadronic states  are parameterized in terms of
the leptonic decay constants, i.e.,
%%buraya kader geldim
%
\begin{eqnarray}\label{Definitions}
\langle0\mid J_{B_c} (0)\mid
B_c(p')\rangle=i\frac{f_{B_c}m_{B_c}^2}{m_{b}+m_{c}},~~~~~~
\langle\chi_{b0}(p)\mid J^\dag_{\chi_{b0}}(0)\mid 0\rangle= -i m_{\chi_{b0}}f_{\chi_{b0}}.
\end{eqnarray}
 Putting all expressions together, the final version  of the
 phenomenological side of the correlation function is obtained as
\begin{eqnarray}\label{PimuPhy}
\Pi^{PHYS}_{\mu}(p^2,p'^2)=\frac{f_{\chi_{b0}}m_{\chi_{b0}}}{(p'^2-m_{B_c}^2)(p^2-m_{\chi_{b0}}^2)}
\frac{f_{B_c}m_{B_c}^2}{m_{b}+m_{c}}\Bigg[f_{1}(q^{2}) P_{\mu}
+f_{2}(q^{2})q_{\mu}\Bigg] +\mbox{...},\nonumber\\
\end{eqnarray}
where we will choose the structures $P_{\mu}$ and $q_{\mu}$, to
evaluate the form factors $f_1$ and $f_2$, respectively.

At QCD side, the correlation function is calculated  in  deep Euclidean region by the
help of the OPE. 
\begin{figure}[h!]
\begin{center}
\includegraphics[width=16cm]{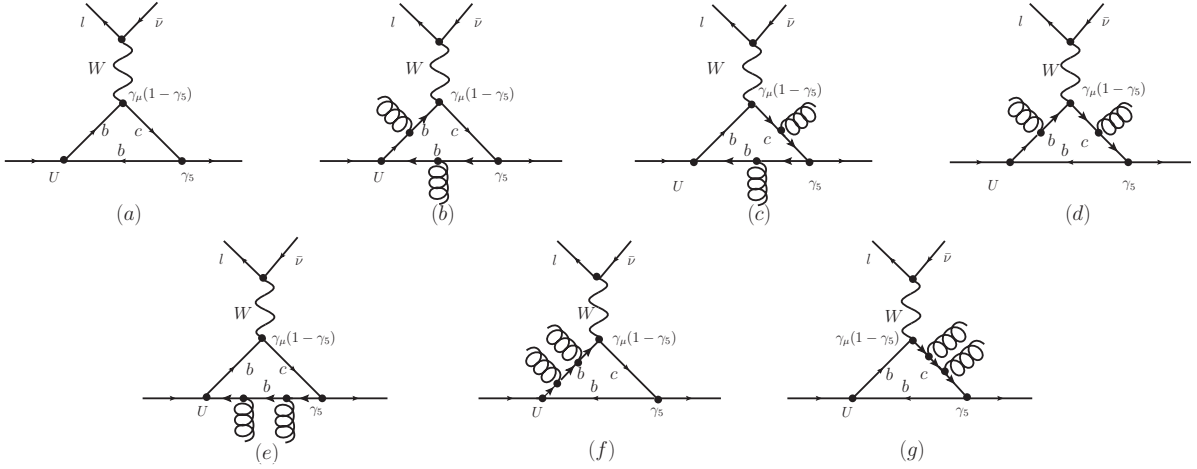}
\end{center}
\caption{Feynman diagrams  contributing to the correlation function
for the $\chi_{b0}\rightarrow B_c\ell\overline{\nu}$ decay:  (a) the bare loop and (b, c, d, e, f, g) two-gluon condensate diagrams.}
\label{Diagrams}
\end{figure}
For this aim, we write the coefficient of each structure in correlation
function  as a sum of a perturbative (diagram a in figure 1) and a
non-perturbative (diagrams b, c, d, e, f and g in figure 1) parts as follows:

\begin{eqnarray}
\Pi^{QCD}_{\mu}&=&(\Pi_1^{pert}+\Pi_1^{nonpert})P_{\mu}+(\Pi^{pert}_2+\Pi^{nonpert}_2)q_{\mu}
\end{eqnarray}
where, the $\Pi_i^{pert}$ functions are written in terms of  double
dispersion integrals in the following way:
\begin{equation}\label{dispersionrelation}
\Pi_i^{pert}=-\frac{1}{(2\pi)^2}\int ds\int
ds'\frac{\rho_{i}(s,s',q^2)}{(s-p^2)(s'-p'^2)}+\textrm{ subtraction
terms},
\end{equation}
where, $\rho_{i}(s,s', q^2)$ are the spectral densities with $i=1$ or $2$. Applying the usual Feynman integral technique to the bare loop
diagram, the spectral densities  are calculated via
Cutkosky rules, i.e., by replacing the quark propagators with Dirac
delta functions: $\frac{1}{p^2-m^2}\rightarrow-2\pi\delta(p^2-m^2)$ implying that all quarks are real. After some calculations, the  spectral densities are obtained as follows:
\begin{eqnarray}\label{specdenrho}
\rho_{1}(s,s',q^2)&=&2N_{c}I_{0}(s,s',q^2)\Big[m_b(m_c-3m_b)-A(h+s)
-B(h+s')\Big],\nonumber\\
\rho_{2}(s,s',q^2)&=&2N_{c}I_{0}(s,s',q^2)\Big[m_b(m_b+m_c)-A(h-s)
+B(h-s')\Big],
\end{eqnarray}
where
\begin{eqnarray}\label{I0AB-coef}
I_{0}(s,s',q^2)&=&\frac{1}{4\lambda^{1/2}(s,s',q^2)},\nonumber\\
A&=&\frac{1}{\lambda(s,s',q^{2})}\Big[(m_b^2-m_c^2)u^{\prime\prime}+us'\Big],\nonumber\\
B&=&\frac{1}{\lambda(s,s',q^{2})}\Big[2(m_b^2-m_c^2)s+su'\Big],
\nonumber\\
h&=&2m_b(m_b-m_c),
\end{eqnarray}
here also $\lambda(s,s',q^{2})=s^{2}+s'^{2}+q^4-2ss'-2sq^{2}-2s'q^{2}$, $u=q^2+s-s'$, $u'=q^2-s+s'$, $u^{\prime\prime}=q^2-s-s'$
and $N_{c}=3$ is the number of colors. The integration region for
the perturbative contribution  in  Eq. (\ref{dispersionrelation}) (bare loop diagram)
is determined requiring that the arguments of the three $\delta$
functions vanish, simultaneously. Therefore, the physical region in
the $s$ and $s'$ plane is described by the following inequality:
\begin{equation}\label{fsu}
-1\leq f(s,s')=\frac{2s[m_{b}^2-m_{c}^2+s'+\frac{u^{\prime\prime}}{2}]}
{\lambda^{1/2}(m_{b}^2,s, m_{b}^2)\lambda^{1/2}(s,s',q^2)}\leq+1.
\end{equation}
For the non-perturbative part, we take into account the two-gluon condensate diagrams (b, c, d, e, f, g) in figure 1. Here we should mention that we deal with the heavy quarks in the 
present work and the heavy quarks' condensates
 are suppressed by inverse  of their masses, so we can ignore them safely. After lengthy calculations for the two-gluon condensates diagrams (b, c, d) correspond to the 
 diagrams with two gluon lines coming out from different quark lines, we get
\begin{equation}\label{NonpertPmu}
 \Pi^{nonpert}_i=\int_0^1 dx \int_0^{1-x} dy  \langle0|\frac{1}{\pi}\alpha_s G^2|0\rangle \Bigg\{\frac{\Theta_i^1}{{\cal D}^5}+\frac{\Theta_i^2}{{\cal D}^4}+\frac{\Theta_i^3}{{\cal D}^3}+
\frac{\Theta_i^4}{{\cal D}^2}\Bigg\},
\end{equation}
where
 \begin{equation}\label{D}
{\cal D}=m_c^2x-m_b^2 r+p'^2xr+p^2yf+x y(p^2+p'^2-q^2),
\end{equation}
and
\begin{eqnarray}\label{Tetas}
\Theta_{1}^{1}&=& \frac{1}{24}\Bigg\{9m_b^5m_cy^2v^2(-1+2y)-18m_b^6y^2v^3-3m_b^4xv\Bigg[3m_c^2\Big(r^2x+2rxy+2wy^2\Big)\nonumber\\
&+& q^2y\Bigg(3r^3x+3r^2ty+ry^2(-14+15x)+y^3(-22+18x+9y)\Bigg)\Bigg]\nonumber\\
&+& 3m_b^3m_cxv \Bigg[m_c^2x\Big(6-19x+13x^2-15y+22xy+12y^2\Big)+q^2y\Bigg(13r^2x^2-6r^2x\nonumber\\
&+&3ry-15ryx+22ryx^2+y^2\Big(15-22y-26x+12x^2+12xy+12y^2\Big)\Bigg)\Bigg]\nonumber\\
&+&m_bm_cq^2x^2yv\Bigg[m_c^2x\left(22-49x+39x^2-62y+66xy+36y^2\right)+q^2y\Bigg(39x^4-88x^3\nonumber\\
&+&66x^3y+3fy\Big(3-7y+6y^2\Big)+x^2\Big(71-128y+36y^2\Big)+xz\Big(11-33y+18y^2\Big)\Bigg)\Bigg]\nonumber\\
&+&3q^4x^3y^2\Bigg[m_c^2\Bigg(f^2+6f^3y+x^3(-11+3x+9y)+fx\Big(10-17x+18xy-26y+18y^2\Big)\Bigg)\nonumber\\
&+&q^2y\Bigg(x^3\Big(31+3x^2-14x+9xy-50y+21y^2\Big)+fx\Big(-20+54y-51y^2+15y^3\nonumber\\
&+&36x-59xy+27xy^2\Big)+f^2(-4+3yz^2)\Bigg)\Bigg]+3m_b^2q^2x^2y\Bigg[q^2y\Bigg(4r^3-2r^3x+16r^2y\nonumber\\
&-&8r^2xy+3ry^2\Big(8-7x+2x^2\Big)+y^3\Big(17-5y-27x+12x^2+6xy\Big)\Bigg)+m_c^2\Bigg(6xy-4x\nonumber\\
&+&5x^2-2x^3-5x^2y+6x^2y^2+f\Big(8y-8xy+12xy^2+6y^2z-1\Big)\Bigg)\Bigg]\Bigg\},\nonumber\\
\Theta_{1}^{2}&=&\frac{1}{96}\Bigg\{9m_b^3m_cv\Bigg[rx^2(-7+13x)+2x^2(-8+11x)y-2y^2\Big(3f+5x-6x^2\Big)+12y^3(f+x)\Bigg]\nonumber\\
&-&9m_b^4v\Bigg[3r^2x^2+6rx^2y+y^2\Big(7-14x+15x^2-16y+18xy+9y^2\Big)\Bigg]+3m_b^2x\nonumber\\
&\times&\Bigg[3m_c^2\Big(6y^4-xr^2+6y^2r^2-2xry+12ry^3\Big)+q^2y\Bigg(8r^2x-11r^2x^2-12ry+54rxy\nonumber\\
&-&38ryx^2-y^2\Big(43+153x^2-48x^3+54ty+23y^2-158x-48xyt-48xy^2\Big)\Bigg)\Bigg]\nonumber\\
&+&m_bm_cxv\Bigg[9m_c^2x\Big(4+13rx-13y+22xy+12y^2\Big)+q^2y\Bigg(312x^4+x^3(-535+528y)\nonumber\\
&+&x^2\Big(313-683y+288y^2\Big)+3y\Big(-9y+53y-88y^2+48y^3\Big)+6x\Big(-9+40y\nonumber\\
&-&58y^2+24y^3\Big)\Bigg)\Bigg]+3q^2x^2y\Bigg[m_c^2\Bigg(r^2\Big(3-29x+24x^2\Big)-y\Bigg(21-x\Big(151-206x+72x^2\Big)\Bigg)\nonumber\\
&+&78y+2xy(-127+72x)+36y^2(-3+4x)+48y^3\Bigg)+q^2y\Bigg(45x^5+x^4(-182+135y)\nonumber\\
&+&x^3\Big(306-584y+315y^2\Big)+x^2\Big(-253+849y-1004y^2+405y^3\Big)+3f\Big(4-30y+69y^2\nonumber\\
&-&59y^3+15y^4\Big)+x\Big(96-500y+965y^2-792y^3+225y^4\Big)\Bigg)\Bigg]\Bigg\},\nonumber\\
\Theta_{1}^{3}&=&\frac{1}{96}\Bigg\{9m_b^2\Bigg[-r^2x^2-2rx^2y+r\Big(1-7xy^2+8x^2y^2\Big)+2y^3(1+8rx)+y^4(-1+8x)\Bigg]\nonumber\\
&+&9m_c^2x\Bigg[r^2x(-3+4x)+2rxy(-5+6x)+2y^2\Big(2-13x+12x^2\Big)+12ty^3+8y^4\Bigg]\nonumber\\
&+&4q^2yx\Bigg[r^2x\Big(17-64x+45x^2\Big)+y\Big(21+x(-154+413x-418x^2+135x^3)\Big)-120y^2\nonumber\\
&+&y^2x(497-718x+315x^2)+3y^3\Big(76-198x+135x^2\Big)+3y^4(-58+75x)+45y^5\Bigg]\nonumber\\
&+&3m_b m_cv\Bigg[52x^4+24xfy^2+12fy^2(-1+2y)+x^3(-61+88y)+3x^2\Big(7-19y+16y^2\Big)\Bigg]\Bigg\},\nonumber\\
\Theta_1^4 &=&\frac{3}{16}\Bigg\{5x^5+y^2f^2(-4+5y)+x^4(-14+15y)+fxy^2(-19+25y)+x^3\Big(13-28y+35y^2\Big)\nonumber\\
&-&x^2\Big(4-13y+48y^2-45y^3\Big)\Bigg\}.
\end{eqnarray}
Here $v=-1+x+y,~ r=-1+x,~ t=-1+2x,~ w=1+x,~ f=-1+y$ and $ ~z=-2+y$. The explicit expressions for $\Theta_{2}^{1,2,3,4}$ correspond to the structure $q_\mu$ are too long, hence we do not present them here.
In a similar way, we calculate the contributions of the diagrams (e, f, g) correspond to two gluon lines coming out from the same quark line. Because of their very lengthy expressions, we do not also depict their 
explicit form here, but we will take into account their contributions in our numerical results.

 The next step is to equate the coefficients of selected structures from both sides in order to get sum rules for the form factors. After applying double
Borel transformations with respect to the variables $p^2$ and $p'^2$
($p^2\rightarrow M^2, p'^2\rightarrow M'^2$)  to
suppress the contributions of the higher states and continuum, the
QCD sum rules for the form factors are obtained as:

\begin{eqnarray}\label{formfactors}
f_{1,2}(q^2)&=&\frac{(m_{b}+m_{c})
}{f_{B_{c}}m_{B_{c}}^2}\frac{1}{f_{\chi_{b0}}m_{\chi_{b0}}}e^{m_{\chi_{b0}}^2/M^{2}}e^{m_{B_{c}}^2/M'^2}\bigg\{-\frac{1}{(2\pi)^2}\int_{4m_{b}^2}^{s_0}ds\int_{(m_b+m_c)^2}^{s_0'}ds'\nonumber \\
&\times&\rho_{1,2}(s,s',q^2)\theta[1-f^{2}(s,s')]e^{-s/M^{2}}e^{-s'/M'^2}+\hat{\cal
B}\Pi_{1,2}^{nonpert}\bigg\},
\end{eqnarray}
where the operator $\hat{\cal
B}$ denotes double Borel transformation.
Note that  to subtract  contributions of the higher states and
continuum, we also apply the quark-hadron duality assumption, i.e., 
\begin{eqnarray}\label{ope}
\rho^{higher states}(s,s') = \rho^{OPE}(s,s') \theta(s-s_0)
\theta(s'-s'_0).
\end{eqnarray}
We also perform the double Borel transformation  as follows:
\begin{itemize}
 \item for the perturbative part, we use 
\begin{equation}\label{Borelrule}
\hat{\cal
B}\frac{1}{[p^2-m^2_1]^m}\frac{1}{[p'^2-m^2_2]^n}\rightarrow(-1)^{m+n}\frac{1}{\Gamma[m]}\frac{1}{\Gamma
[n]}e^{-m_{1}^2/M^{2}}e^{-m_{2}^2/M'^2}\frac{1}{(M^{2})^{m-1}(M'^2)^{n-1}}.
\end{equation}
\item For the non-perturbative part, first we make the transformation
\begin{equation}
 \hat{\cal
B}\frac{1}{[p^2-f(p^{'2})]^n}=(-1)^n \frac{e^{-f(p^{'2})/M^2}}{\Gamma[n](M)^{n-1}},
\end{equation}
to write  the terms containing $p^{'2}$ in exponential form. Then we apply the following rule to transform the ($p^{'2}\rightarrow M'^2$): 
\begin{equation}
 \hat{\cal
B} e^{-\alpha p^{'2} }=\delta(\frac{1}{M'^2}-\alpha),
\end{equation}
where $\alpha$ is a function of quarks' masses as well as the parameters used in Feynman parametrization.
\end{itemize}

\section{Numerical Results}

In this section, we numerically analyze the related form factors, obtain their fit function in terms of $q^2$ and estimate the decay width of the channel under consideration. 
In calculations, we use the input parameters as presented in table I. 
\begin{table}[ht]
\centering
\rowcolors{1}{lightgray}{white}
\begin{tabular}{cc}
\hline \hline
   Parameters  &  Values    
           \\
\hline \hline
$m_{c}$              & $(1.275\pm0.015)~ GeV$\\
$m_{b}$              & $(4.7\pm0.1)~ GeV$\\
$ m_e $              &   $ 0.00051  $ $GeV$ \\
$ m_\mu $            &   $ 0.1056   $ $GeV$ \\
$ m_\tau $           &   $ 1.776 $  $GeV$ \\
$ m_{\chi_{b0}}$    &   $ (9859.44\pm0.42\pm0.31) $ $MeV$   \\
$ m_{B_c} $      &   $ (6.277\pm0.006) $ $GeV$   \\
$ f_{B_c} $      &   $(400\pm 40) ~MeV $   \\
$ f_{\chi_{b0}} $      &   $(175\pm55) ~MeV $   \\
$ G_{F} $            &   $ 1.17\times 10^{-5} $ $GeV^{-2}$ \\
$  V_{cb} $ &   $ (41.2\pm1.1)\times 10^{-3} $   \\
$ \langle0|\frac{1}{\pi}\alpha_s G^2|0\rangle$        &   $ (0.012\pm0.004)$ $GeV^{4}$   \\
 \hline \hline
 
\end{tabular}
\caption{Input parameters used in our calculations \cite{Nakamura,E Veliev,cbquarkmass,Wang}.}
\end{table}

The sum rules  for the form factors denote
that they also depend on four auxiliary parameters, namely continuum
thresholds $s_{0}$ and $s'_{0} $ and Borel mass parameters $M^2$ and
$M'^2$.  The continuum thresholds are not
completely arbitrary but they are in correlation with the energy of the
excited state in initial and final channels. Considering this point and the fact that the result of the physical quantities (form factors)
should weakly depend on these
parameters, we choose the intervals $s_{0}
=(97.7-99.2)~ GeV^2$ and $s_{0}' =(40-41)~GeV^2 $ slightly higher than the mass of pole squared of the initial and final mesonic channels for the continuum thresholds.  The Borel
parameters $M^2$ and $M'^2$ also are not physical quantities, hence the form
factors should be independent of them. The reliable regions for the
Borel parameters $M^2$ and $M'^2$ can be determined by requiring that
not only the contributions of the higher states and continuum are
effectively suppressed, but   contribution of the operators with
the higher dimensions are small, i.e., the sum rules for form factors converge. As a result of these
requirements, the working regions for these parameters are determined to be $15~ GeV^2
\leq M^2\leq30~ GeV^2$ and $10~ GeV^2\leq M'^2\leq20 ~GeV^2$. 
\begin{figure}[h!]
\begin{center}
\includegraphics[totalheight=6cm,width=7cm]{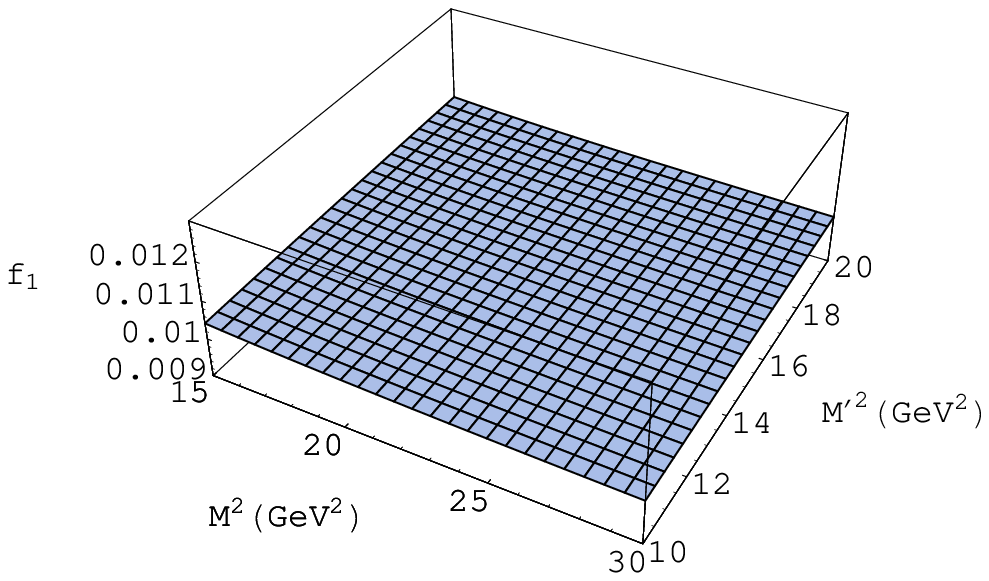}
\includegraphics[totalheight=6cm,width=7cm]{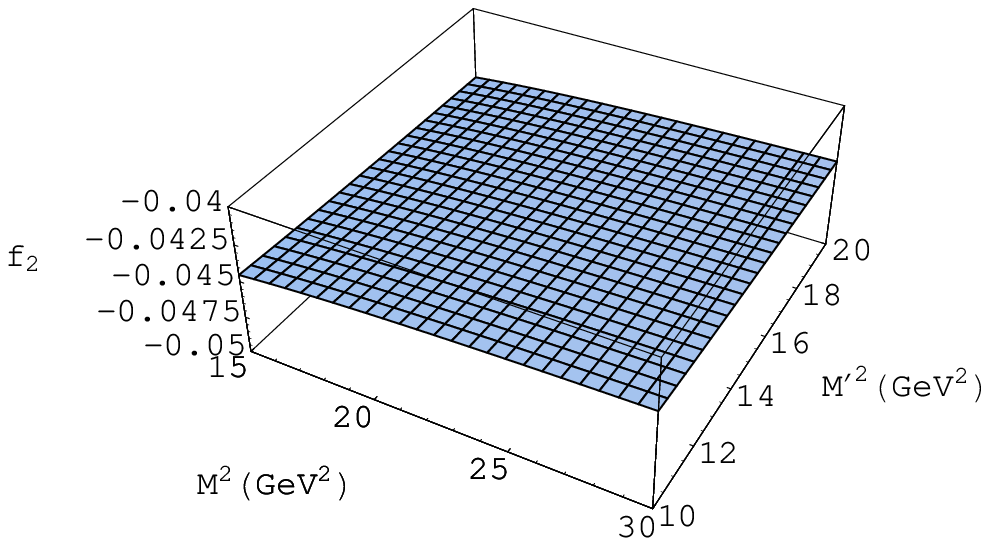}
\end{center}
\caption{Dependence of the form factors $f_1$ and $f_2$
 on Borel mass parameters $M^{2}$ and $M'^{2}$ at $q^2=1~ GeV^2$.}
\label{f1f2MsqMpsq3Dgraph}
\end{figure}
The
dependence of form factors $f_1$ and $f_2$ on Borel masses at $q^2 =
1~GeV^2$ are plotted in figure 2. From this figure, we see good stability of the form factors with respect to the variations of the Borel mass parameters at their working regions.
 To see the convergence of the OPE, we compare both perturbative and non-perturbative contributions to the form factors in figure 3 at $q^2 =1~GeV^2$ and the presented Borel windows.
\begin{figure}[h!]
\begin{center}
\includegraphics[totalheight=6cm,width=7cm]{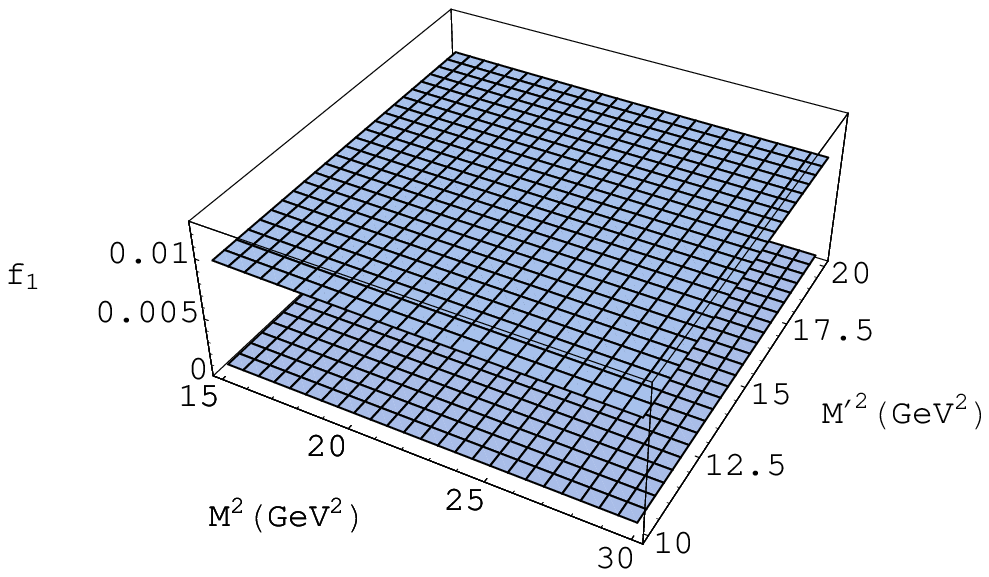}
\includegraphics[totalheight=6cm,width=7cm]{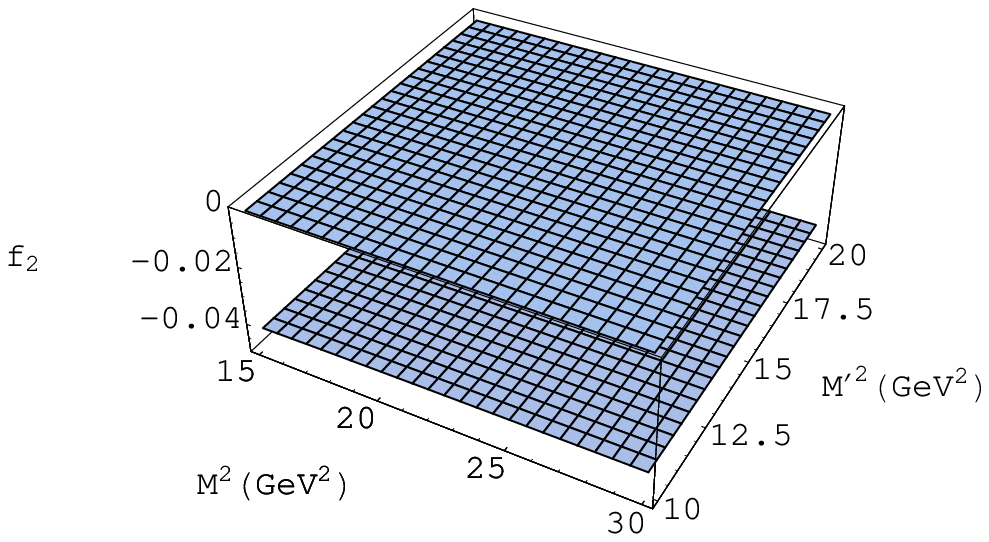}
\end{center}
\caption{ Comparison between perturbative and non-perturbative contributions to the form factors at $q^2=1~ GeV^2$ and chosen Borel windows. The upper (lower) plane belongs to the perturbative contribution
in  $f_1$ ($f_2$).}
\label{f1f2MsqMpsq3Dgraph}
\end{figure}
From this figure and our numerical calculations, it is found that the ratio of non-perturbative contribution to that of perturbative is $0.08$ and $0.05$ for $f_1$ and $f_2$, respectively.  Hence, the 
non-perturbative contribution constitutes only $7.5$\% and $4.8$\%  of the total results respectively for the form factors $f_1$ and $f_2$. This means that the series of sum rules for the form factors are
 convergent. In the presented Borel windows, the contributions of
the excited and continuum states are  exponentially suppressed. This guarantees the reliability of the sum rules and isolation of the ground state from the excited states and continuum.

Our calculations show that the form factors are
truncated at  $q^2 \simeq9~GeV^2$ (see figure 4).  After this point up to the higher limit of the $q^2$, the sum rules predictions are not reliable (for details see for instance \cite{braun1,braun2}).  However, we need their fit functions  in the whole physical region, $ m_{l}^2
\leq q^2 \leq (m_{\chi_{b0}}-m_{B_{c}})^2$ to estimate the decay width of the $\chi_{b0}\rightarrow
B_{c}\ell\overline{\nu}$ transition.  To extend our results to the
full physical region, we search for parameterization of the form
factors in such a way that in the region $0 \leq q^2 \leq 9~
GeV^2$, predictions of this parameterization coincide with the sum rules
results. The following parametrization adjust well  the $q^2$ dependence of the form factors:
\begin{equation}\label{fitfunction}
f_{i}(q^2)=\frac{a}{(1- \frac{q^{2}}{m_{fit}^{2}})}+\frac{b}{(1-
\frac{q^{2}}{m_{fit}^{2}})^{2}},
\end{equation}
where, the values of the parameters $a$, $b$ and $m_{fit}$ obtained using $M^2=25~GeV^2$ and $M^{'2}=15~GeV^2$ for the
$\chi_{b0}\rightarrow B_{c}\ell\overline{\nu}$ channel are given in  table
II.
\begin{table}[h]
\centering
\begin{tabular}{|c|c|c|c|} \hline
& a  & b & $m_{fit}^2~(GeV^2)$\\\cline{1-4}
 $f_{1}(\chi_{b0}\rightarrow B_{c}\ell\overline{\nu})$ & -0.055 &0.062 & 21.86\\\cline{1-4}
 $f_{2}(\chi_{b0}\rightarrow B_{c}\ell\overline{\nu})$ & 0.225 &-0.254 & 19.79\\\cline{1-4}
 \end{tabular}
 \vspace{0.5cm}
\caption{Parameters appearing in the fit function of the form factors. \label{tab:1}}
\end{table}

\begin{figure}[h!]
\begin{center}
\includegraphics[totalheight=6cm,width=7cm]{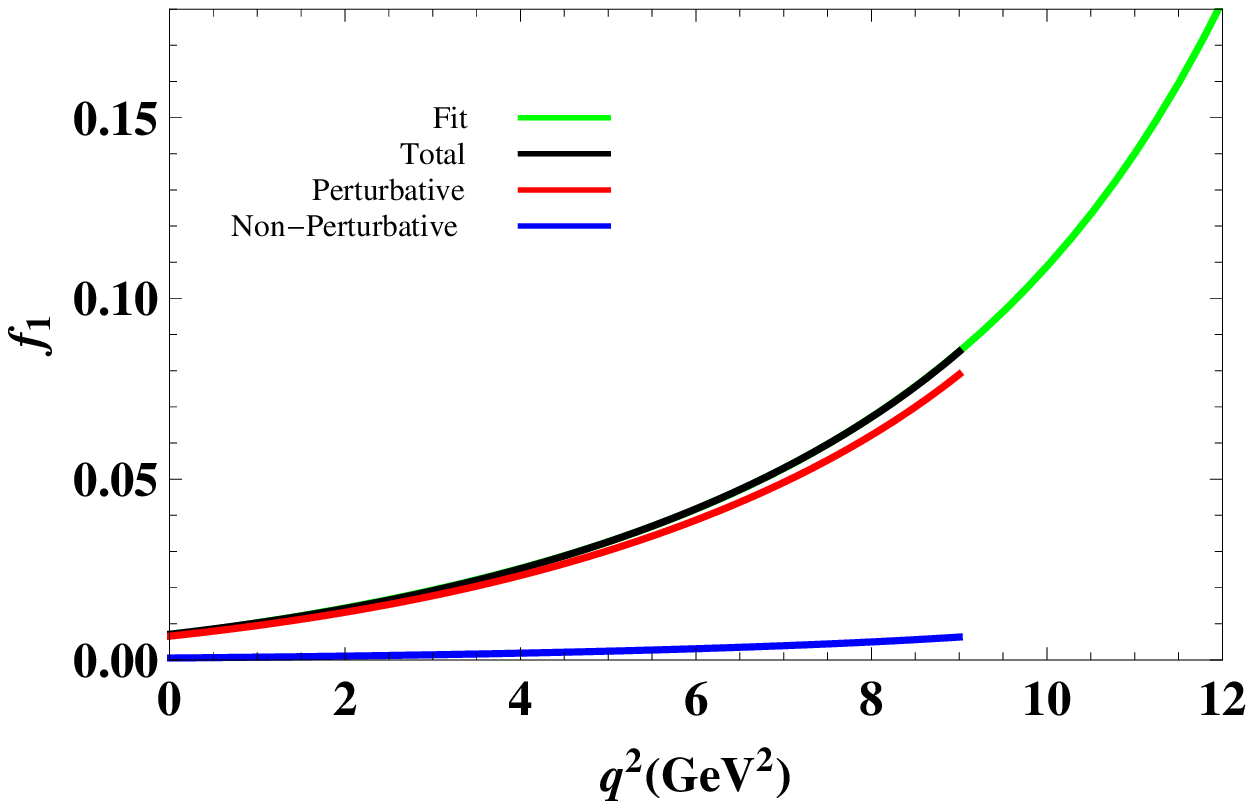}
\includegraphics[totalheight=6cm,width=7cm]{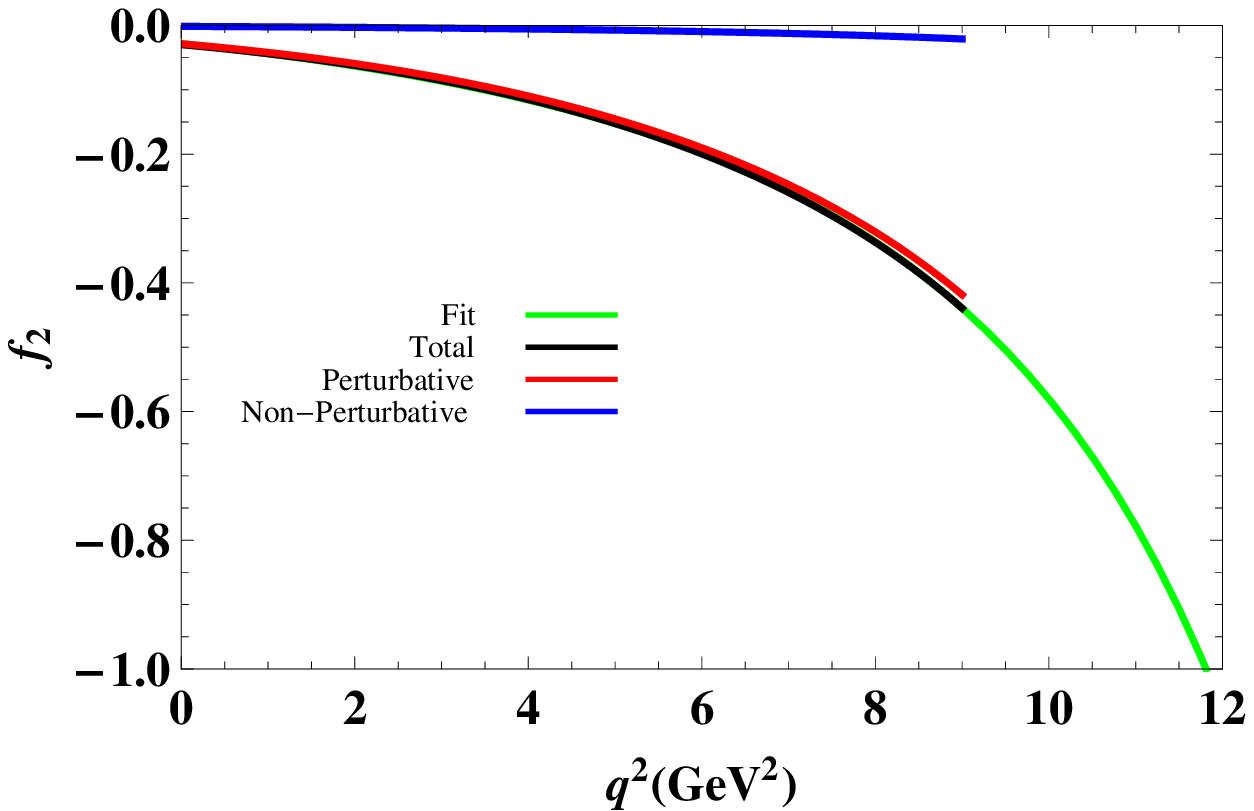}
\end{center}
\caption{Dependence of the form factors $f_1$ and $f_2$ on $q^{2}$ at $M^2=25~GeV^2$ and $M^{'2}=15~GeV^2$.} \label{f1fitandf2fit}
\end{figure}

We  depict the dependence of
form factors $f_1$ and $f_2$ on $q^2$  obtained directly from the sum rules as well as the fit parametrization at whole physical region in figure \ref{f1fitandf2fit}. In the case of
sum rules predictions, we present the perturbative, non-perturbative and total contributions in this figure.

Having obtained the behavior of the form factors in terms of $q^2$ at whole physical region,  we would like to calculate the 
decay width  of the process under consideration. Using the amplitude previously discussed, the differential decay
width  for $\chi_{b0}\rightarrow B_{c}\ell\overline{\nu}$ is
obtained in terms of form factors as:
\begin{eqnarray}\label{Decaywidth}
\frac{d\Gamma}{dq^2}&=&\frac{G_{F}^2}{192\pi^{3}m_{B_{c}}^{3}}
|V_{cb}|^2\lambda^{1/2}(m_{B_{c}}^{2},m_{S}^{2},q^{2})
\left(\frac{q^{2}-m_{\ell}^{2}}{q^{2}}\right)^{2}
\Bigg\{\frac{q^{2}}{2}\Bigg[|f_{1}(q^{2})|^{2}
(2m_{B_{c}}^{2}+2m_{S}^{2}-q^{2})
 \nonumber \\
&+&2(m_{B_{c}}^{2}-m_{S}^{2})Re[f_{1}(q^{2})f_{2}^{*}(q^{2})]+
|f_{2}(q^{2})|^{2}q^{2}\Bigg]-\frac{(q^{2}+m_{\ell}^{2})}{2q^{2}}\Bigg[|f_{1}(q^{2})|^{2}(m_{B_{c}}^{2}-
m_{S}^{2})^{2}
 \nonumber \\
&+&2(m_{B_{c}}^{2}-m_{S}^{2})q^{2}Re[f_{1}(q^{2})f_{2}^{*}(q^{2})]+
|f_{2}(q^{2})|^{2}q^{4}\Bigg] \Bigg\}.
\end{eqnarray}
Performing
integration over $q^2$  in Eq.
(\ref{Decaywidth}) in the interval $m_{l}^2\leq
q^2\leq(m_{\chi_{b0}}-m_{B_{c}})^2$,  we obtain the expression for the total decay width. The numerical values of the decay width at different lepton channels are  presented in Table III.
\begin{table}[h]
\centering
\begin{tabular}{|c||c|}\hline
   &  $\Gamma(GeV)$  \\ \hline \hline
$\chi_{b0}\rightarrow B_c e\overline{\nu}_{e}$ & $1.46\times
10^{-14}$ \\\hline
  $\chi_{b0}\rightarrow B_c \mu\overline{\nu}_{\mu}$ & $1.45\times 10^{-14}$ \\\hline
  $\chi_{b0}\rightarrow B_c \tau\overline{\nu}_{\tau}$ & $0.91\times 10^{-14}$ \\\hline
  \hline
\end{tabular}
 \vspace{0.5cm}
\caption{Numerical results for  decay rate at different lepton channels.}
\label{Tabledecaywidth}
\end{table}
The errors in the values of the decay rates in table III
are due to uncertainties in determination of the working regions for continuum thresholds and Borel mass parameters as well as
errors of the other input parameters.

\section{Conclusion}
In the present work, we studied the semileptonic $\chi_{b0}\rightarrow
B_{c}\ell\overline{\nu}, (\ell=(e,\mu,\tau))$ decay channel within the framework of the three-point QCD sum rules. In particular, taking into account the two-gluon diagrams as non-perturbative contributions, we obtained
the QCD sum rules for the form factors entered the transition matrix elements. After obtaining the working regions for the auxiliary parameters, we found the behavior of the form factors in terms of  $q^2$ in whole
physical region. The fit function of the form factors were then used to estimate the decay rates at different lepton channels.
Any measurement on the form factors as well as decay rate of the channel under consideration and comparison of the obtained results with theoretical predictions in the present study can give valuable information 
about the internal structures of the participating mesons specially  nature of the scalar  $\chi_{b0}(1P)$ state.

\section{Acknowledgment}
This work is supported in part by Scientific and Technological
Research Council of Turkey (TUBITAK) under project No: 110T284 and
partly by Kocaeli University Scientific Research Center (BAP)  under
project No: 2011/52.

\end{document}